\documentclass[twoside,10pt,a4paper]{newFNLstyle}
\usepackage{graphics}
\usepackage{cite}
\usepackage[dvips]{graphicx}
\usepackage{amssymb}
\usepackage{amsmath}
\begin{document}

\volnumpagesyear{0}{0}{000--000}{2001}
\dates{received date}{revised date}{accepted date}

\title{CAT'S DILEMMA -- TRANSITIVITY VS\mbox{.} INTRANSITIVITY}
\authorstwo{EDWARD W.~PIOTROWSKI* and MARCIN MAKOWSKI**}
\affiliationtwo{Institute of Mathematics, University of Bia\l
ystok,\\  Lipowa 41, PL-15424 Bia\l ystok, Poland}
\mailingtwo{ep@alpha.uwb.edu.pl*, mmakowski@alpha.pl**}


\maketitle

\markboth{Cat's Dilemma -- transitivity vs\mbox{.} intransivity}{Piotrowski and Makowski}

\pagestyle{myheadings}


\begin{abstract}
  We study a simple example of a sequential game
  illustrating  problems connected with making  rational
  decisions that are universal for social sciences. The set of
  chooser's optimal decisions  that manifest
  his preferences in case of a constant strategy of the adversary (the offering
player),
  is investigated. It turns out that the order imposed
  by the player's rational preferences can be intransitive.
  The presented quantitative results imply a revision of the
  "common sense" opinions stating that preferences showing
  intransitivity are paradoxical and undesired.
\end{abstract}

\section{Introduction}

The intransitivity can occur in  games with three or more
strategies if the strategies  $A$, $B$, $C$ are such that $A$
prevails over $B$, $B$ prevails over $C$, and  $C$ prevails over
$A$ ($A>B>C>A$). The most known example of intransitivity is the
children game "Rock, Scissors, Paper" ($R,S,P$) where $R>S>P>R$.
The other interesting example of intransitive order is the
so-called Condorcet's paradox, known since  XVIIIth century.
Considerations regarding this paradox led Arrow in the XXth
century to prove the theorem stating that there is no procedure of
successful choice that would meet the democratic
assumptions\cite{r1}. The importance of this result to mathematical political
science is comparable to G\"{o}del's Incompleteness Theorem in logic
\cite{r2}.
\newline It seems logical to choose an order, in a consistent way
between things we like. But what we prefer often depends on how
the choice is being offered \cite{r8,r7}. This paradox was
perceived by many researchers and analysts  (for instance Stan Ulam described this
in his book "Adventures of a Mathematician", some problems with intransitive 
options can be found in  \cite{r9,r10}). On the other hand
scientists have a penchant for classifications
(rankings) on basis of linear orders and this (we think) follows
from such intransitive preferences there are so suspicious
for many researchers.

        In the paper, we  present  quantitative analysis of a model,
which can be illustrated by the Pitts's experiments with cats,
mentioned in the Steinhaus diary\cite{r4} (Pitts noticed that a
cat facing choice between fish, meat and milk prefers fish to
meat,  meat to milk, and milk to fish!). This model finds its
reflection in the \emph{principle of least action} that controls
our mental and physical processes, formulated by Ernest Mach
\cite{r3} and referring to Ockham's razor principle.\newline
Pitts's cat, thanks to the above-mentioned food preferences,
provided itself with a balanced diet. In our work, using elementary
tools of linear algebra, we obtained the relationship between the optimal cat's
strategy and frequencies of appearance of food pairs. Experiments with rats
confirmed Pitts's observations.
Therefore, it is interesting to
investigate  whether intransitivity of preferences will provide a
balanced diet also in a wider sense  in more or less abstract
situations involving decisions. Maybe in the class of randomized
behaviors we will find the more effective ways of nutrition? The
following sections constitute an  attempt at providing
quantitative answer to these questions. The analysis of an
elementary class of models of making  optimal decision presented
below permits only determined behaviors, that is such for which
the agent must make the choice.

        Through this analysis we wish to contribute to dissemination of
theoretical quantitative studies of nondeterministic algorithms
of behaviors which are essential for economics and
sociology -- this type of analysis is not in common use.
The geometrical interpretation presented in this article can turn out very
helpful in understanding of various stochastic models in use.

\section{Nondeterministic cat}
\begin{figure}[htbp]
   \centering{
 \includegraphics[
    height=2in,
    width=2.1in]%
   {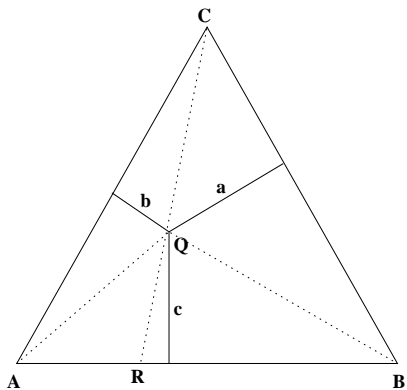}}\caption{The barycentric coordinates.}\label{os1}
\end{figure}

\begin{figure}[htbp]
         \centering{
       \includegraphics[
          height=2in,
          width=2.1in]%
         {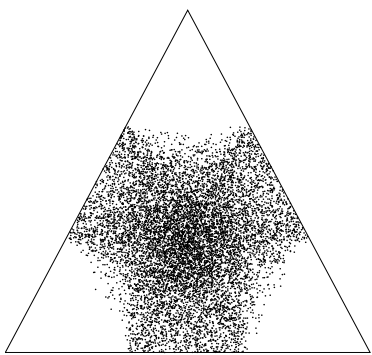}}\caption{Image of the three-dimensional cube on simplex.}
         \label{hex}
\end{figure}

\begin{figure}[htbp]
          \centering{
        \includegraphics[
           height=1.5in,
           width=2in]
          {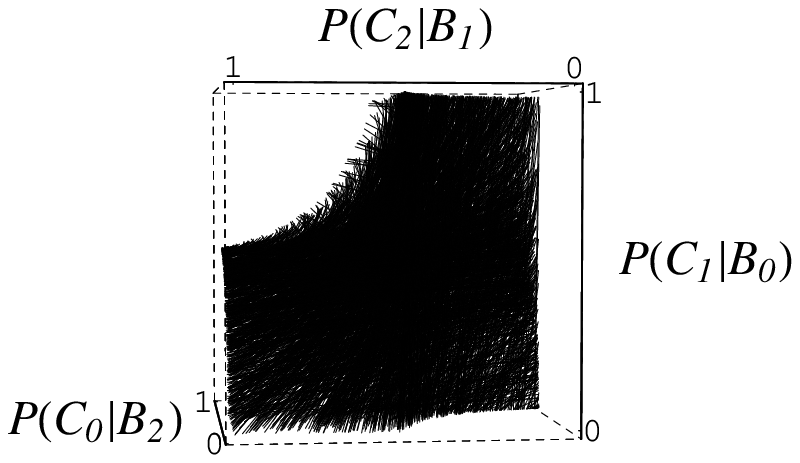}
         \includegraphics[
           height=1.5in,
           width=2
           in]%
       {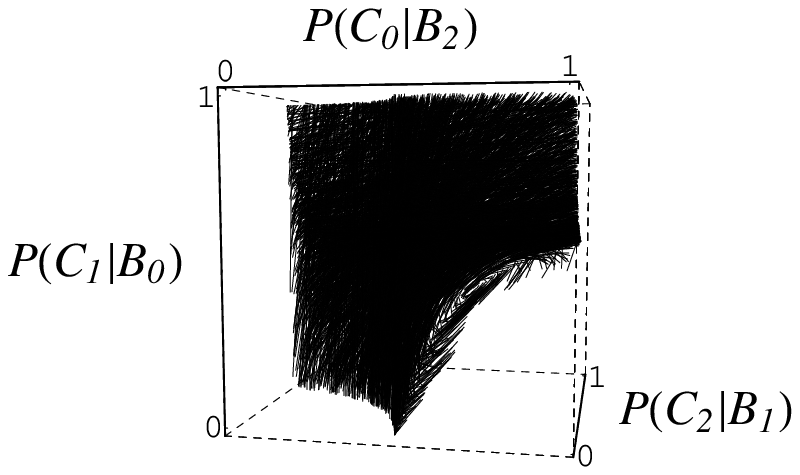}\\\includegraphics[
       height=1.5in,
       width=2in]%
       {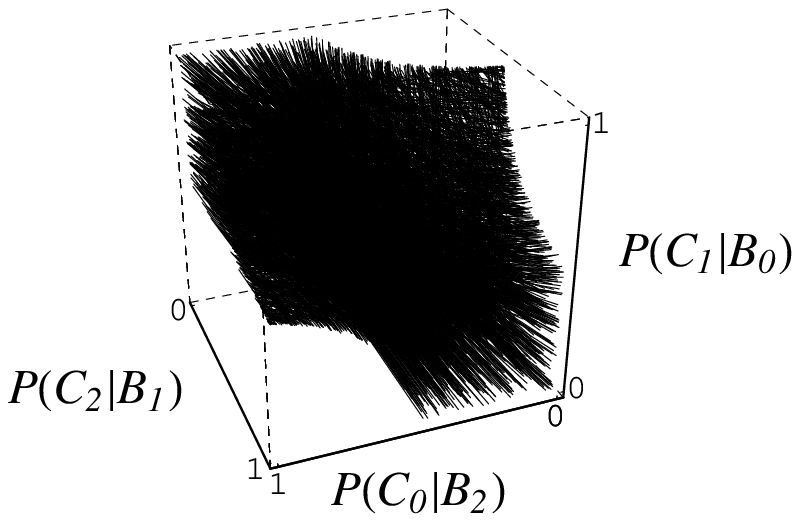}\includegraphics[
              height=1.5in,
              width=2
              in]%
          {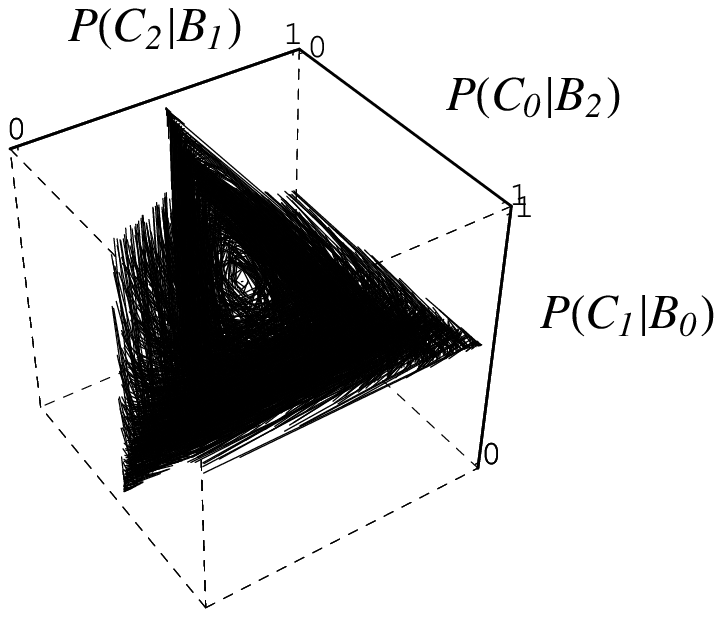}}
\caption{The inverse image of area of frequencies $(q_0,q_1,q_2)$
          that enable realization of the optimal strategy, see Appendix A.} 
\label{cube}
\end{figure}

Let us assume that a cat is offered three types of food (no.~1,
no.~2 and no.~3), every time in pairs of two types, whereas the food
portions are equally attractive regarding the calories, and each
one has some unique components that are  necessary for the cat's
good health. The cat knows (it is accustomed to) the
frequency of occurrence of every pair of food and his strategy
depends on only this frequency. Let us also assume that the cat cannot consume both
offered types of food at the same moment, and that it will never
refrain from making the choice. The eight (${2^{3}}$) possible
deterministic choice functions  ${f_{k}}$:
\begin{equation}
f_k:\{(1,0), (2,0),(2,1)\}\to \{0,1,2\},\qquad k=0,\ldots,7
\end{equation}
are defined in  Table 1.
\begin{table}[htbp]
\caption{The table defining all possible choice functions
$f_k$.}\vspace{1ex} \centering\footnotesize
\begin{tabular}{|c|c|c|c|c|c|c|c|c|} \hline
\vphantom{$\int^1$} function \vphantom{$F^K$}$f_k$: &
${f_{0}}$ & ${f_{1}}$ & ${f_{2}}$ & ${f_{3}}$ & ${f_{4}}$ &
${f_{5}}$ & ${f_{6}}$ & ${f_{7}}$\\[1pt]
\hline
\vphantom{$\int^1$}$f_{k}(1,0)=$\vphantom{$F^K$} & 0 & 0 & 0 & 0 & 1 & 1 & 1 & 1
\\[1pt]\hline \vphantom{$\int^1$}$f_{k}(2,0)=$\vphantom{$F^K$}
 & 0 & 0 & 2 & 2 & 0 & 0 & 2 & 2 \\[1pt]\hline\vphantom{$\int^1$}
$f_{k}(2,1)=$ \vphantom{$F^K$}& 1 & 2 & 1 & 2 & 1 & 2 & 1 & 2
\\[1pt]\hline \vphantom{$\int^1$}frequency $p_k$: & $p_0$ & $p_1$ & $p_2$ & $p_3$ & $p_4$
& $p_5$ & $p_6$ & $p_7$
\\[1pt]\hline
\end{tabular}
\end{table}
\noindent The functions $f_{2}$  and $f_{5}$ determine
intransitive orders. The parameters $p_k$, $k=0,\ldots,7$ give the
frequencies of appearance of the choice function  in the
nondeterministic algorithm (strategy) of the cat
($\sum_{k=0}^{7}p_k = 1$, $p_k\geq 0$ for $k=0,\ldots,7$).\newline
We will show the relationship between the frequency of occurrence of individual
type of food
in cat's diet and the frequencies of occurrence of food pairs.\newline
Let us denote the frequency of occurrence of the pair $(k,j)$  as
$q_m$, where $m$ is the number of food that does not occur in the
pair $(k,j)$ ($\sum_{m=0}^{2}q_m = 1$). This  denotation causes no
uncertainty because there are only three types of food. When the
choice methods $f_k$ are selected nondeterministically, with the
respective intensities $p_k$, the frequency $\omega_m$, $m=0,1,2$,
of occurrence of individual food in cat's diet are according to
Table 1\mbox{.} given as follows:
\begin{itemlist}
\item food no.~0: $\omega_0=(p_0+p_1+p_2+p_3)q_2+(p_0+p_1+p_4+p_5)q_1$,
\item food no.~1: $\omega_1=(p_4+p_5+p_6+p_7)q_2+(p_0+p_2+p_4+p_6)q_0$,
\item food no.~2: $\omega_2=(p_2+p_3+p_6+p_7)q_1+(p_1+p_3+p_5+p_7)q_0$.
\end{itemlist}
Three equalities above can be explained with the help of the
conditional probability concept. Let us denote $B_{3-(j+k)}=\{(j,k)\}$,
$P(B_{j}) = q_{j}$ and $C_{j}=\{j\}$ for 
$j,k=0,1,2$, $ j\neq k$.
The number $P(C_{k} | B_{j})$
indicates the probability of choosing the food of number $k$, when
the offered food pair does not contain the food of number $j$.
Since 
the events of choosing different pairs of  food are
disjoint and comprise all the space of elementary events. Hence,
for each  food chosen, we have the following relation:
\begin{equation} 
\omega_k
=P(C_k)=\sum_{j=0}^{2}P(C_{k}|B_{j})P(B_{j}),\,\, k=0,1,2.
\end{equation}
By inspection of the table of
the functions $f_k$, $k\negthinspace=\negthinspace0,\dots,7$, we easily get the following relations:
\begin{eqnarray}\label{par}
P(C_{0}|B_{2})=P(\sum_{k=0}^{7} f_{k}(B_2)=0)=p_0+p_1+p_2+p_3\,,\nonumber\\
P(C_{0}|B_{1})=P(\sum_{k=0}^{7} f_{k}(B_1)=0)=p_0+p_1+p_4+p_5\,,\nonumber\\
P(C_{1}|B_{0})=P(\prod_{k=0}^{7} f_{k}(B_0)=1)=p_0+p_2+p_4+p_6\,,\\
P(C_{1}|B_{2})=P(\prod_{k=0}^{7} f_{k}(B_2)=1)=p_4+p_5+p_6+p_7\,,\nonumber\\\nonumber
P(C_{2}|B_{1})=P(\prod_{k=0}^{7} \frac{f_{k}(B_1)}{2}=1)=p_2+p_3+p_6+p_7\,,\\\nonumber
P(C_{2}|B_{0})=P(\prod_{k=0}^{7} \frac{f_{k}(B_0)}{2}=1)=p_1+p_3+p_5+p_7\,,\\\nonumber
\end{eqnarray}
and $P(C_{0}|B_{0})=P(C_{1}|B_{1})=P(C_{2}|B_{2})=0$.\newline
Frequency of the least preferred food, that is the function
$\min(\omega_0,\omega_1,\omega_2)$, determines the degree of
the diet completeness. Since
$\omega_0+\omega_1+\omega_2=1$, the most valuable way of choosing
the food by the cat occurs for such probabilities
$p_0,\ldots,p_7$, that the function
$\min(\omega_0,\omega_1,\omega_2)$  has the maximal value, that is
for
 \begin{equation}
 \label{maximum}
 \omega_0=\omega_1=\omega_2=\tfrac{1}{3}.
 \end{equation}
 Any vector $\vec{p}=(p_0,\ldots,p_7)$
 (or six conditional probabilities $(P(C_1|B_0),$ $P(C_2|B_0)$,\newline$
  P(C_0|B_1)$,$P(C_2|B_1)$,$P(C_0|B_2),P(C_1|B_2))$),
 which for a fixed triple $(q_0,q_1,q_2)$ fulfills
 the system of equations (\ref{maximum})
 will be called an cat's \emph{optimal strategy}\/.\newline
 Let us study this strategy in more details and subject it
 to geometrical analysis.
 For given $q_0$, $q_1$, $q_2$ all optimal strategies are calculated. 
 The system of equations (\ref{maximum})
 has the following matrix form:
\begin{equation}\label{matrix maximal}
\left( \begin{array}{ccc} P(C_0|B_2) & P(C_0|B_1) & 0 \\ P(C_1|B_2) & 0 &
P(C_1|B_0) \\ 0 & P(C_2|B_1) & P(C_2|B_0)
\end{array} \right)
\left( \begin{array}{ccc} q_2 \\ q_1 \\ q_0
\end{array} \right)=\tfrac{1}{3}
\left( \begin{array}{ccc} 1 \\ 1 \\ 1
\end{array} \right),
\end{equation}
and its solution:
\begin{eqnarray}\label{solution}
q_2&=&\tfrac{1}{d}\bigg(\frac{P(C_0|B_1)+P(C_1|B_0)}{3}-P(C_0|B_1)P(C_1|B_0)\negthinspace\bigg),\nonumber\\
q_1&=&\tfrac{1}{d}\bigg(\frac{P(C_0|B_2)+P(C_2|B_0)}{3}-P(C_0|B_2)P(C_2|B_0)\negthinspace\bigg),\\
q_0&=&\tfrac{1}{d}\bigg(\frac{P(C_1|B_2)+P(C_2|B_1)}{3}-P(C_1|B_2)P(C_2|B_1)\negthinspace\bigg),\nonumber
 \end{eqnarray}
 defines a mapping of the three-dimensional cube $[0,1]^3$
 in the space of parameters $(P(C_0|B_2),P(C_0|B_1),P(C_1|B_0))$ into a
triangle
 in the space of parameters $(q_0,q_1,q_2)$, where
 $d$ is the determinant of the matrix of
parameters $P(C_j|B_i)$. The barycentric coordinates \cite{r6} of a
point of this triangle are interpreted as
 the probabilities $q_0$, $q_1$ and $q_2$.
 These numbers represent the heights $a$, $b$ and $c$
 or the areas $P_{QAB}$, $P_{QBC}$ and $P_{QAC}$ of three smaller triangles
 determined by the point $Q$ (cf.~Fig.~\ref{os1}), or the lengths
 of the segments formed by the edges of the triangle by
 cutting them with the straight lines passing through the point $Q$  and the
 opposite vertex of the triangle.
  Hence e.g. $\frac{\normalsize q_1}{q_2}=\frac{a}{b}=\frac{P_{QBC}}{P_{QAC}}
  =\frac{|RB|}{|RA|}$, where the symbol $|RB|$ represents
  length of the segment.
  \newline  The next picture (Fig.~\ref{hex}) presents
  the image of the three-dimensional cube in this simplex.
  It determines the area of frequency $q_m$
  of appearance of individual choice alternatives between
  two types of food in the simplex, for which the optimal strategy exists.
  In order to present the range of the nonlinear representation
  of our interest, the authors illustrated it with the values of
  this representation for 10,000 randomly selected points with respect
  to constant probability distribution on the cube.
  Justification of such equipartition of probability
  may be found in Laplace's principle of insufficient reason\cite{r5}.
  In our randomized model the a priori probability of the
  fact that the sum of probabilities $P(C_j|B_k)$  is smaller than a
  given number $\alpha\in [0,1]$ equals $\alpha$. The absence of optimal
  solutions outside the hexagon forming the shaded part of the picture (Fig.~2) is obvious,
  since the bright (non-dotted) part of the picture represents
  the areas, for which $q_0>\frac{1}{3}$
  (or $q_1>\frac{1}{3}$, or $q_2>\frac{1}{3}$), and the total
  frequency of appearance of pairs $(0,1)$ or $(0,2)$ must be
   at least $\frac{1}{3}$ in order to assure the
  completeness of the diet with respect of the ingredient $0$
  (but this cannot happen because when $q_0>\frac{2}{3}$, then
  $q_1+q_2=1-q_0<\frac{1}{3})$.\newline
  The system of equations (\ref{matrix maximal}) can be transformed into the
following form:
 \begin{equation}
 \left( \begin{array}{ccc}
 q_2 & -q_1 & 0 \\[2pt]
 -q_2 & 0 & q_0 \\[2pt]
 0 & q_1 & -q_0
 \end{array} \right)
 \left( \begin{array}{ccc}
 P(C_0|B_2)\\[2pt] P(C_2|B_1)\\[2pt] P(C_1|B_0)
 \end{array} \right)\,=\,
 \left( \begin{array}{ccc}
 \frac{1}{3}-q_1 \\[2pt] \frac{1}{3}-q_2 \\[2pt] \frac{1}{3}-q_0
 \end{array} \right),
 \end{equation}
 which allows to write out the inverse transformation
   to the mapping defined by
  equations (\ref{solution}). By introducing the parameter $\lambda$
  we may write them as follows:
  \begin{equation}
  P(C_0|B_2)=\frac{\lambda}{3q_2}\,,\,\,
  P(C_2|B_1)=\frac{\lambda-1+3q_1}{3q_1}\,,\,\, P(C_1|B_0)=\frac{\lambda+1-3q_2}{3q_0}\,.\label{solu2}
  \end{equation}
  A whole segment on the unit cube
  corresponds to one point of the simplex,
  parameterized by $\lambda$. The range of this representation
  should be limited to the unit cube, which gives the
  following conditions for the above subsequent equations:
  \begin{equation}
  \lambda\in[0,3q_2], \,\,\,\,\lambda\in[1-3q_1,1],\,\,\,\, \lambda\in[3q_2-1,2-3q_1].
  \end{equation}
  The permitted values of the parameter $\lambda$\,
  form the common part of these segments, hence it is nonempty for:
  \begin{equation}
  \max(0,1-3q_1, 3q_2-1)\,\,\leq\,\, \min(2-3q_1,3q_2,1).
  \end{equation}
  Therefore
  \begin{equation}
  \lambda\in[ \max(0,1-3q_1, 3q_2-1),\min(2-3q_1,3q_2,1)].
  \end{equation}
   It may be now noticed that for any triple
  of probabilities belonging to the hexagon,
  there exists an optimal solution within a
  set of parameters ($(P(C_0|B_2),P(C_0|B_1)$,\\$P(C_1|B_0))$). If we assume the equal
  measure for each set of frequencies of occurrence
  of food pairs as the triangle point, then we may
  state that we deal with optimal strategies in $\frac{2}{3}$ of
  all the cases (it is the ratio of area of regular
  hexagon inscribed into a equilateral triangle).
  The inverse image of the area of frequencies $(q_0,q_1,q_2)$ of food
  pairs that enable realization of the optimal strategies,
  which is situated on the cube of all possible strategies,
  is presented by four consecutive plots in Fig.~\ref{cube}. We present
  there the same configuration observed from different
  points of view. The segments  on the figures
  correspond to single points of the frequency triangle of the
  individual food pairs. The greatest concentration of the segments
  is observed in two areas of the cube that correspond to
   intransitive strategies\footnote{See section 4.}. The bright area in the center
  of the cube, which may be seen in the last picture, belongs to the effective strategies
  -- effective in
  the subset of frequencies of a small measure $(q_0,q_1,q_2)$ of the food
  pairs appearance. Among them, the totally
  incidental behavior is located, which gives
  consideration in equal amounts to all the mechanisms of deterministic choice
$p_j=p_k=\frac{1}{8}\,$.

\section{Example of an optimal strategy}
  The formulas (\ref{solu2}) that map the triangle
  into a cube can be used to find an
  optimal strategy in  cases, when the
  probabilities $(q_0,q_1,q_2)$  of appearance of individual
  pairs of the products are known.
  Let us assume that $q_0=\frac{1}{2}$\,, $q_1=\frac{1}{3}$ and $q_2=\frac{1}{6}$\,.
  Then, according to the formulas (\ref{solu2}),
  we have $P(C_1|B_0)=\frac{1}{3}+\frac{2\lambda}{3}$\,, $P(C_0|B_2)=2\lambda$\,,
$P(C_2|B_1)=\lambda$\,,
 where $\lambda\negthinspace\in\negthinspace[0,\frac{1}{2}]$. Selecting
 $\lambda\negthinspace=\negthinspace\frac{1}{4}$
  we have: $P(C_0|B_2)=\frac{1}{2}$\,, $P(C_2|B_1)=\frac{1}{4}$\,, $ P(C_1|B_0)=\frac{1}{2}$\,.
  We may now show the solution of equations (\ref{par}),
  e.g.: $p_0=\frac{1}{2}$, $p_5=p_7=\frac{1}{4}$
  and $p_j=0$ for others parameters.
  We will obtain the following frequencies of occurrence
  of individual foods in the diet:
 \begin{eqnarray}\nonumber
 \omega_0=(p_0+p_5)q_1+p_0q_2=\frac{1}{4}+\frac{1}{12}=\frac{1}{3}\,,\\
 \omega_1=p_0q_0+(p_5+p_7)q_2=\frac{1}{4}+\frac{1}{12}=\frac{1}{3}\,,\\\nonumber
 \omega_2=(p_5+p_7)q_0+p_7q_1=\frac{1}{4}+\frac{1}{12}=\frac{1}{3}\,.
 \end{eqnarray}
  The above calculations of the frequency $\omega_j$  confirm
  optimality of the indeterministic algorithm determined in this example.

 \section{Intransitive nondeterministic decisions}
\begin{figure}[htbp]\label{star}
          \centering{
        \includegraphics[
           height=2in,
           width=2.1in]
          {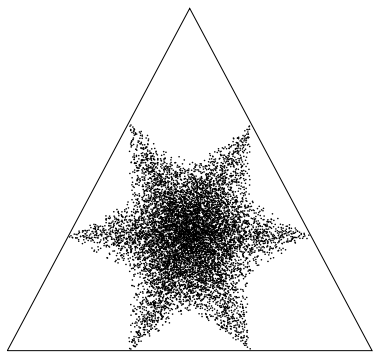}}\caption{Optimal intransitive strategies.}
\end{figure}

\begin{figure}[htbp]
           \centering{
         \includegraphics[
            height=2in,
            width=2.1in]%
           {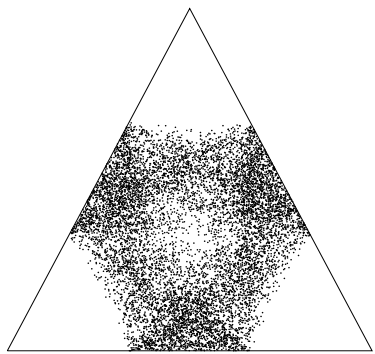}}\caption{Optimal transitive strategies.}\label{trans}
\end{figure}

 In the case of  random selections we may talk about order relation
 \emph{food no.~0} $<$  \emph{food no.~1}, when from the offered pair $(0,1)$
 we are willing to choose the food no.~$1$  more often than the food no.~$0$
 ($P(C_0|B_2)<P(C_1|B_2)$). Therefore we have two intransitive orders:
 \begin{itemize}
 \item $P(C_0|B_2)<\frac{1}{2}$, $P(C_2|B_1)<\frac{1}{2}$, $P(C_1|B_0)<\frac{1}{2}$\,.
 \item $P(C_0|B_2)>\frac{1}{2}$, $P(C_2|B_1)>\frac{1}{2}$, $P(C_1|B_0)>\frac{1}{2}$\,.
 \end{itemize}
 It is interesting to see in which part of the simplex
 of parameters $(q_0,q_1,q_2)$ we may take optimal intransitive
 strategies. They form the six-armed star composed of two triangles,
 each of them corresponding to one of two possible intransitive orders (Fig.~\ref{star}).
 They dominate in the central part of triangle, near point
$q_0=q_1=q_2=\frac{1}{3}$.
 They form darkened part of area inside the star. Optimal transitive
strategies cover the same area of the simplex
 as all  optimal strategies, however they occur less
 often in the center of the simplex. We illustrated this
 situation in the next picture (Fig.~\ref{trans}). In areas of high
 concentration of optimal transitive strategies,
 one of three frequencies  $q_0$, $q_1$, $q_2$  looses its significance -- two
 from three pairs of the food occur with considerable predominance.
  We have enough information to be able to compare
  the applicability range of different types
  of optimal strategies. Let us assume the same measure
  of the possibility of occurrence of determined proportion
  of all three food pairs. This assumption means
  that the probability of appearance of the situation determined
  by a point  in the triangle-domain of parameters $(q_0,q_1,q_2)$
  does not depend on those parameters.
  Two thirds of strategies  are optimal.
   There are 33\%\footnote{They are measured by the area of
equilateral triangle inscribed into a regular hexagon.} of
circumstances, which allow for the use of the optimal strategies
that belong to the specified
  intransitive order. There are 44\% ($\frac{4}{9}$) of situations of any
  order that favor optimal strategies, what follows from the
  fact that they are measured by the surface of regular star,
  and its area is equal to double area of the triangle corresponding
  to one intransitive order reduced by the area of the hexagon
  inscribed into the star. So we have:
  $\frac{1}{3}+\frac{1}{3}-\frac{2}{9}=  \frac{4}{9}$. Appearance of
  the number $\frac{2}{9}$  in the calculation can be easily explained
  by the observation that the area of the regular six-armed star is
   two times bigger than the area of the hexagon inscribed into it.
  This number (22\%) is the measure of the events that favor
  both types of intransitive strategies.

  It is worth
 to stress that in the situation that favors  optimal
 strategies we can always find the strategy that determines
 the transitive order (see Fig.~\ref{trans}). However, we should remember that this
 feature concerns only the simple model of the cat's
 behavior, and does not have to be true in the cases of more
 complicated reaction mechanisms.
\section{Conclusions}
In this article, we used a stochastic variant of \emph{the
principle of least action}\/. Perhaps dissemination of usage of this principle
will lead to formulation of many interesting conclusions and observations.\newline
We presented a method, which allows
successful analysis of intransitive orders that still are
surprisingly suspicious for many researchers.
More profound analysis of this phenomenon can have importance everywhere
where the problem of choice behavior is studied. For instance in
economics (description of the customer preference toward products- marketing strategy)
or in political science where  the problem of voting exists. Analysis of intransitive
orders is a serious challenge to those who seek description of our reasoning process.

        The quantitative observations from the previous section show that
intransitivity, as the way of making the decision, can provide the
diet completeness for the cat from our example. Moreover, the
intransitive optimal strategies constitute the major part of all
optimal strategies. Therefore, it would be wrong to prematurely
acknowledge the preferences showing the intransitivity as
undesired. Perhaps there are
  situations, when only the intransitive orders
 allow obtaining the optimal effects. The most intriguing problem that remains open,
 is to answer the question whether
 there exists a useful
 model of optimal behaviors, which gives
 the intransitive orders, and for which it would
 be impossible to specify the transitive optimal strategy
 of identical action results. Showing the impossibility of
 building such  constructions would cause marginalization of
 the practical meaning of intransitive orders. On the other hand,
indication of this type of models would force us to accept the
intransitive ordering.
 \section*{Acknowledgements}
The authors are grateful to prof.~Zbigniew Hasiewicz and prof.~Jan
S\l adkowski for useful discussions and comments.
\appendix 
\section{}
 The following mini-program written in the language {\em Mathematica 5.0}\/ 
generate four plots in Fig.~\ref{cube}.

\[
\begin{split}
In[1]:=~~&
c = N[\tfrac{4}{3\sqrt{3}}];\\[2ex]
&gencorrect := 
 Module[\{a = c \,\{ Random[], Random[]\}\},
         While[\,a[[2]] > \tfrac{2}{3},\\
&\phantom{gencorrect := Modul} a = c\, \{ 
Random[],  Random[]\}\,];\, a\,];\\[2ex]
&setpoint := 
      Module[\{q1, a, x, y, v = 1, w = 1\}, 
        While[\,\tfrac{2}{c}\, v + \tfrac{3}{2}\, w > 1, 
         \\
&\phantom{setpoint :=  Modul} a = gencorrect;\, q1 = a[[2]]; 
    \,  x = a[[1]] - \tfrac{c}{2};\\ 
&\phantom{setpoint :=  Modul}  y = a[[2]] - \tfrac{1}{3};\, v = Abs[x]; 
         \, w = Abs[y]\,]; \\
&\phantom{setpoint :=  M}\{q1, N[\tfrac{\sqrt{3}}{2} \,x - 
\tfrac{1}{2} \,y +
 \tfrac{1}{3}]\}\,];\\[2ex]
&setsegment := 
      Module[\{a, q1 = 1, q2 = 1, l1 = 1, l2 = 0\},\\
&\phantom{setsegment := Mo}While[l1 > l2\,, 
          a = setpoint; q1 = a[[1]]; 
          q2 = a[[2]]; \\
 &\phantom{setsegment := MoWhi} l1 = 
            Max[0, 1 - 3\, q1, 3\, q2 - 1];\\ 
 &\phantom{setsegment := MoWhi} l2 = Min[2 - 3\, q1, 3\, q2, 
              1]\,]; \\
&\phantom{setsegment := Mo}\{\{\tfrac{l1}{3\, q2}, \tfrac{1 - l1}{3\, q1}, 
\tfrac{1 - 3\, q2  + l1}{3 (1 - q1 - q2)}\}, 
\{\tfrac{l2}{3\, q2}, \tfrac{1 - l2}{3\, q1}, 
\tfrac{1 - 3\, q2  + l2}{3 (1 - q1 - q2)}\}\}\,];\\[2ex] 
&fig[x\_\,, y\_\,, z\_] := 
      Show[Graphics3D[\{GrayLevel[ .0], 
Thickness[ .002],\\
&\phantom{fig[x\_\,, y\_\,, z\_]:=Show[Gr} 
            Table[Line[setsegment], \{2000\}]\}],\\
 &\phantom{fig[x\_\,, y\_\,, z\_]:=Sh}ViewPoint \rightarrow \{x, y, z\}, 
        \,Axes \rightarrow True, \\
&\phantom{fig[x\_\,, y\_\,, z\_]:=Sh}
        AxesLabel \rightarrow \{ \text{"\negthinspace 
  P($C_0\mid B_2$)"},  \text{"\negthinspace P($C_2\mid B_1$)"}, 
 \text{"\negthinspace   P($C_1\mid B_0$)"}
\}, \\
&\phantom{fig[x\_\,, y\_\,, z\_]:=Sh}
Ticks \rightarrow\{\{0, 1\}, \{0, 1\}, \{0, 1\}\}, \\
&\phantom{fig[x\_\,, y\_\,, z\_]:=Sh}
        BoxStyle \rightarrow Dashing[\{ .02,  .02\}], \\
&\phantom{fig[x\_\,, y\_\,, z\_]:=Sh}
        AxesStyle \rightarrow 
          Thickness[ .005]\,];\\
&fig[-6,  .7,  .3]\\
&fig[.6,  -3,  .3]\\
&fig[1.3, 3.4, 2]\\
&fig[3,-2, 3]
\end{split}
\]\\

\begin{thebibliography}{10}

\bibitem{r1}
K.~J.~Arrow, {\em Social Choice and Individual Values}\/,
 Yale University Press, New York (1951).

\bibitem{r2}
M.~Gardner, {\em Time Travel and Mathematical Bewilderments}\/, Freeman, New York (1988).


\bibitem{r8}
A.~Tversky,  {\em Elimination by aspects: A theory of choice}\/,
Psychological Review, {\bf 79} (1972) 281-299.

\bibitem{r7}
A.~Tversky,  {\em Intransitivity of Preferences}\/, Psychological Review,
{\bf 76} (1969) 31-48.

\bibitem{r9}
J.~Y.~Halpern, {\em Intransitivity and Vagueness}\/, Principles of Knowledge Representation
and Reasoning, Proceedings of the Ninth International Conf., Whistler, Canada (2004); arXiv:cs/0410049.

\bibitem{r10}
E.~Groes, H.~J.~Jacobsen, T.~Tranas,
{\em Testing the Intransitivity Explanation of the Allais paradox},  Theory and Decision {\bf 47} (1999), 229-245.

\bibitem{r4}
H.~Steinhaus, {\em Memoirs and Notes}\/ (in Polish), Aneks, London
(1992).

\bibitem{r3}
E.~Mach, {\em The Science of Mechanics}\/, Open Court, LaSalle, IL (1960).

\bibitem{r6}
 {\em Encyclopaedia of mathematics on cd-rom}\/, Kluwer Academic Publishers,
 Dordrecht (1997).

\bibitem{r5}
P.~Dupont, {\em Laplace and the Indifference Principle in the
'Essai philosophique des probabilités'}\/, Rend\mbox{.}
Sem\mbox{.} Mat\mbox{.} Univ\mbox{.} Politec\mbox{.} Torino, {\bf
36} (1977/78) 125-137.
\end{thebibliography}

\end{document}